\documentclass[12pt]{article}

\usepackage{graphicx}
\usepackage{amsmath}

\newcommand{\eventa}{\mathcal{F}}
\newcommand{\samps}{\Omega}

\author{Rom\`an Zapatrin\thanks{Dept. of Informatics, The State Russian Museum, In\.zenernaya 4, 191186, St.Petersburg, Russia; e-mail: Roman.Zapatrin at gmail.com}}

\title{Quantum contextuality in classical information retrieval}

\begin{document}

\maketitle

\begin{abstract}
Document ranking based on probabilistic evaluations of relevance
is known to exhibit non-classical correlations, which may be
explained by admitting a complex structure of the event space,
namely, by assuming the events to emerge from multiple sample
spaces. The structure of event space formed by overlapping sample
spaces is known in quantum mechanics, they may exhibit some
counter-intuitive features, called quantum contextuality. In this
Note I observe that from the structural point of view quantum
contextuality looks similar to personalization of information
retrieval scenarios. Along these lines, Knowledge Revision is
treated as operationalistic measurement and a way to quantify the
rate of personalization of Information Retrieval scenarios is
suggested.
\end{abstract}

\section{The evolution of information needs}\label{sinfoneeds}

The notion of information needs was clearly formulated by Tailor
\cite{taylor1962}. Along with the development of IR systems the
very structure of information needs, of queries was subject to
evolution. Briefly, its mainstream can be described as a
transition (read upwards)
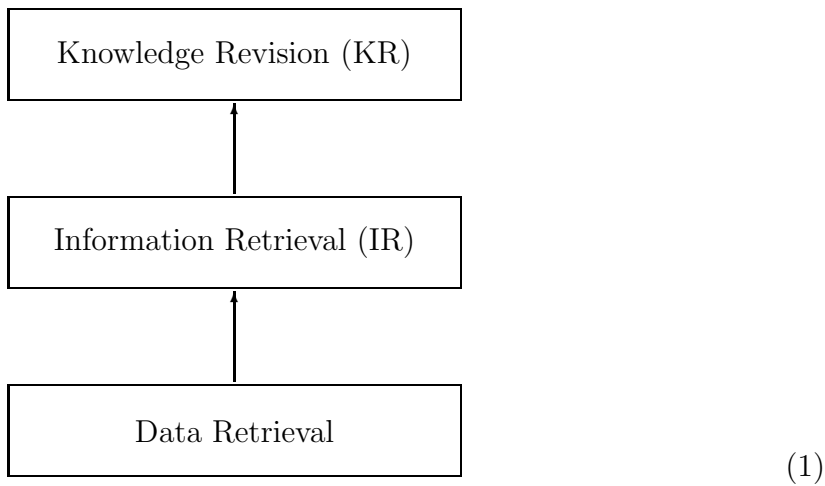
\begin{equation}\label{eretrevol}
\unitlength1mm
\begin{picture}(80,66)
\put(0,50){\framebox(60,12){\mbox{Knowledge Revision (KR)}}}
\put(30,37.5){\vector(0,1){12}}
\put(0,25){\framebox(60,12){\mbox{Information Retrieval (IR)}}}
\put(30,12.5){\vector(0,1){12}}
\put(0,0){\framebox(60,12){\mbox{Data Retrieval}}}
\end{picture}
\end{equation}
each stage using the previous one as a background. Information
Retrieval uses Data Retrieval environment yet modifying the
structure of queries, as formulated by Lancaster ``An information
retrieval system does not inform (i.e. change the knowledge of)
the user on the subject of his inquiry. It merely informs on the
existence (or non-existence) and whereabouts of documents relating
to his request'' \cite{lancaster1968}. Then the next stage is the
increasing personalization of search. The user interacts with an
IR environment having a goal to update the state of his knowledge
(belief) rather than to retrieve a particular document. This way
Information Retrieval serves for Knowledge Revision (KR).

How quantum mechanics comes? The chain \eqref{eretrevol} can be
compared with the transition from classical mechanics, dealing
with the absolute character of the values measured, to quantum
mechanics, where the result of a measurement is a result of an act
of will of an observer rather than retrieving a pre-existing
value. In both extreme cases, the retrieval act is nothing but a
measurement. Similar to the evolution of the notion of
measurement, the retrieval metaphors evolve.

We shall deal and with the general notion of Information Needs
(IN), ranging them in four levels \cite{taylor1962}
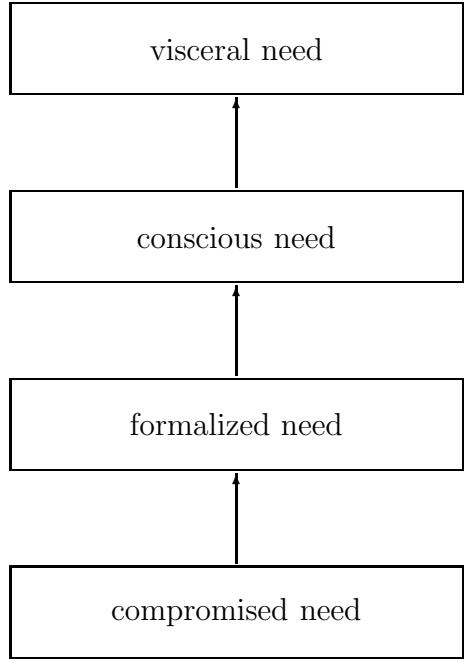
\begin{equation}\label{einlevels}
\unitlength1mm
\begin{picture}(80,86)
\put(0,75){\framebox(60,12){\mbox{visceral need}}}
\put(30,62.5){\vector(0,1){12}}
\put(0,50){\framebox(60,12){\mbox{conscious need}}}
\put(30,37.5){\vector(0,1){12}}
\put(0,25){\framebox(60,12){\mbox{formalized need}}}
\put(30,12.5){\vector(0,1){12}}
\put(0,0){\framebox(60,12){\mbox{compromised need}}}
\end{picture}
\end{equation}
with the following meaning
\begin{itemize}
    \item The visceral need is the actual, but unexpressed, need for information.
    \item The conscious need is a within-brain description of the need.
    \item The formalized need is a formal statement of the question.
    \item The compromised need is the question as presented to the information system.
\end{itemize}
The chain \eqref{eretrevol} reflects the upwards transition in the
above list, and the personalization tightly approaches to the
visceral IN. In this Note I deal with the quantification of
personalization -- the crucial part of Knowledge Revision -- using
quantum metaphor. The technical basis for this quantitative
approach is formed of the following research lines:

\begin{itemize}
    \item Simulation of quantum contextuality effects by finite
    automata and the evaluation of the amount of memory required
    for this simulation \cite{memorycost2011}. Our basic idea is to
    revert this argumentation and to evaluate the features of a
    quantum system, which can be in certain sense simulated by
    giver IR environment.
    \item The evaluations of violations of classical propbabilistic
    laws by index term probabilities, carried out by
    Melucci \cite{melucci2012} and the quantitative evaluation of
    the amount of contextuality by Svozil \cite{svozil2011}
\end{itemize}

\section{On the nature of non-classical
correlations}\label{snoncl}

In general, non-classical correlations appear when Kolmogorovian
probability model is no longer applicable. The basic point of Kolmogorovian
model is the existence of a (single) sample space $\samps$. The
events are subsets of $\samps$, while the points of the sample
space are elementary and independent.

In order to test this or that model, we employ Accardi's
statistical invariants \cite{accardi2006}, they allow to test the
applicability of Kolmogorovian model. Given:
\begin{itemize}
    \item a family of discrete maximal observables
$\{A_\alpha : \alpha=1,\ldots T\}$ ($T$ being finite), each
observable $A_\alpha$ takes the finite number of values
$a^{(\alpha)}_{j
\alpha}$ labelled by $j_\alpha=1,\ldots,n$
    \item the experimentally measurable conditional probabilities
$p_{j_\alpha,j_\beta}(\beta\mid\alpha)$
\begin{equation}\label{eacc32}
    p_{j_\alpha,j_\beta}(\beta\mid\alpha)
    \;=\;
    P\left(A_\beta = a^{(\beta)}_{j \beta}\right\rvert
    \left.A_\alpha = a^{(\alpha)}_{j \alpha}\right)
\end{equation}
\end{itemize}
The problem is: does there exist a probability space $(\samps
;\eventa ; P)$ and $T$ measurable partitions $A_{j}^{(\alpha)}$ of
cardinality $n$ (the number of distinct values of each observable
is assumed to be the same)
\begin{equation*}\label{eaccaa}
    A_{j}^{(\alpha)}, \alpha=1,\ldots T,\,j=1,\ldots n
\end{equation*}
such that for any $\alpha,\beta=1,\ldots T$ one has
\begin{equation}\label{eacckolm}
    P\left(A^{(\beta)}=a_{j}^{(\beta)} \mid A^{(\alpha)}=a_{i}^{(\alpha)}\right)
    \;=\;
    \frac{P\left(A_{j}^{(\beta)}\cup A_{i}^{(\alpha)}\right)}{P\left(A_{j}^{(\beta)}\right)}
\end{equation}
In order to get the answer, a linear programming problem is to be
solved \cite{acfe1981}, that is, the problem of the existence of a
single sample space is finitely decidable.

\medskip

\noindent In the sequel we shall need the special case of three observables
$A,B,C$, each taking only two values $a_1,a_2$ for $A$, $b_1,b_2$
for $B$ and $c_1,c_2$ for the observable $C$. The transition
probability matrices for each pair of observables, being
bistochastic, each has only one numeric parameter, denote the
appropriate matrices as
\begin{equation}\label{ebistoch}
    \begin{array}{l@{\;=\;}c@{\;=\;}c}
      P(A\mid B)&P&\left(%
\begin{array}{cc}
  p & 1-p \\
  1-p & p \\
\end{array}%
\right) \\
      P(B\mid C)&Q&\left(%
\begin{array}{cc}
  q & 1-q \\
  1-q & q \\
\end{array}%
\right) \\
      P(C\mid A)&R&\left(%
\begin{array}{cc}
  r & 1-r \\
  1-r & r \\
\end{array}%
\right) \\
    \end{array}
\end{equation}
then these transition probabilities can be described by a
Kolmogorovian model (that is, they are produced by a single sample
space) if and only if
\begin{equation}\label{eaccardi}
    \lvert p+q-1\rvert
    \leq r \leq
    1-\lvert p-q \rvert
\end{equation}

\section{The operationalistic metaphor}\label{sopermeth}

There is a straightforward analogy between IR and the process of
measurement. There is a search machine, which we may treat to be
prepared in certain state, and there is an observer, which
performs a measurement. It is typical that the preparation of
query system does not assume a query asked by the user, this
causes a mismatch, which is to be handled.

The situation when a mismatch between the preparation and
measurement occurs is a source of paradoxes and counter-intuitive
observable consequences of quantum mechanics. It results in the
possible randomness of single accounts, though previous stages
were deterministically prepared. To deal with it, context
translation is introduced as handling the mismatch between state
preparation and measurement. In quantum mechanics this
metaphorically looks as follows \cite{svozil2004}. Suppose an
electron is prepared, using Stern-Gerlach device, in pure spin
stat along $z$ axis, always showing spin up. Then we decide to ask
the so-prepared electron a complementary question: ``what is
direction of spin along the $x$ axis?'' Quantum mechanics tells us
that the electron is completely incapable to store more than one
bit of information (assuming this is not so leads to direct
experimental contradictions). That is why the electron gives a
random reply on this query. This is what makes it different from
deterministic query agents, who are not able to handle improper
input, on which they offer no answer.

Modern IR environments are no longer so rigid, they easily handle
any kind of input: if you ask them, almost always you get an
answer, but sometimes the relevance of this answer for you
personally may be of zero value. To overcome this, search engines
are configured to track user's requests, or, in other words, to
keep the context associated with particular user and his present
role. Altogether, each such particular action I call knowledge
revision scenario. In practice this is done by seeding pebbles
along the way the user goes through the jungles of World Wide Web,
say, by storing browser's cookies. These pebbles are, after all,
just sequences of bits. Now suppose our task is to judge to what
extent the act of measurement is personalized, let us view it from
a perspective of quantum measurement. To do it, recall a series of
recent works summarized in \cite{memorycost2011}.

\section{Quantifying the personality in Knowledge Revision scenarios}\label{smemorycost}

In brief, quantum contextuality manifests itself as follows: when
measuring quantum systems, the result may depend on which other
compatible observables are measured simultaneously. Furthermore,
these other observables may be just intended to be measured rather
than really measured. This cloud of potentially co-measurable
values is referred to as \textbf{context}. When simulating a
quantum system by agents with internal memory (recall that, as
told above, quantum system are so smart that they behave in this
way without having internal memory), the agent will attain
different internal states in course of carrying out a sequence of
elementary queries. The minimal amount of memory needed to
simulate particular manifestations of quantum contextuality is
called memory cost of this quantum effect. The paper
\cite{memorycost2011} explores the memory cost of simulating quantum
contextuality effect observed on singlet states of positronium. It
gives a clue to draw a correspondence:
\[\mbox{quantum contextuality}\;\longrightarrow\;
\mbox{its memory cost}
\]
In general, the memory cost increases as more and more
contextuality constraints are considered. The complexity of
contextuality constrains depends, in its turn, on the dimension of
the state space of the system in question.

I suggest the following technical idea. The argumentation of the
authors of \cite{memorycost2011} is reverted. We start with an IR
environment and ask how complex quantum contextual features it may
exhibit? Furthermore, we may reduce the answer to just a number
(or a string of numbers), namely, the dimensionality (or a tensor
product structure -- TPS \cite{zanardi2001}) of a quantum system
demonstrating similar context dependence. \[\mbox{KR
scenario}\;\longrightarrow\;
\mbox{quantum system}
\]
How to do this? What is to be simulated? Here, I dwell only on the
logical and certain probabilistic aspects of simulation. To do it,
the proper tools to deal with the structure of the collection of
properties of a system are introduced.

\paragraph{Overlapping contexts.}\label{sgreechie} It was observed
by different authors that complex IR systems are
not well described by probabilistic models based on a single
sample space. In \cite{robertson2004} it was explicitly shown that
Bayesian reasoning in its direct for fails and, in order to get
adequate evaluations, when writing conditional probabilities $P(A
\mid B)$ one should take care about specifying the context -- a
particular sample space, in which these conditional probabilities
are calculated. In the meantime, the small sample spaces are not
separated - thy overlap, there are events belonging to different
contexts. It occurs that the classical contingency table

\medskip

\begin{tabular}{|l|c|c|c|}
  \hline
   & RETRIEVED & NOT RETRIEVED &  \\
  \hline
  RELEVANT & $A\cap B$ & $A\cap \bar{B}$ & $A$ \\
  NON-RELEVANT & $\bar{A}\cap B$ & $\bar{A}\cap \bar{B}$ & $\bar{A}$ \\
  \hline
   & $B$ & $\bar{B}$ &  \\
  \hline
\end{tabular}

\medskip

\noindent ceases to be adequate. The reason is that even within a single
scenario both $A$ and $B$ may belong to different contexts, in
particular, $\bar{A}$ is no longer uniquely defined by $A$ (the
same to $B$ and $\bar{B}$). How to capture this structure? A tool
of combinatorial nature is needed to describe overlapping
contexts. First note that a single sample space is structureless,
all its elements are equally (un)related with each other. In case
of overlapping contexts this is no longer the case. A graphical
(and combinatorial) way to capture such relations was suggested by
R.Greechie (see \cite{greechie-gudder} for an overview). The idea
is to \begin{itemize}
    \item[(i)] consider all the elements of all sample
spaces together
    \item[(ii)] label each element with a tag pointing
to appropriate context
\end{itemize}
The Kolmogorovian probabilities (and hence Bayesian inference)
come from the fact that the logic of statements about the
appropriate sample space is Boolean. In case of pasted contexts
this is no longer so, the structure of all the statements about
the IR environment is no longer Boolean.

\medskip

\noindent How contextuality effects come? Mainly, in the form of
Kochen-Specker reasoning stating that particular hypothetical
probability assignments do not exist such as a total probability
distribution on the whole diagram viewed as a single sample space.
The consequence of such results is signaling that the evaluation
of conditional probabilities based on standard Bayes model will be
no loner adequate. For examples of such violations in quantum
mechanics see \cite{memorycost2011}, in IR this also takes place,
see, for instance \cite{melucci2012}. Quantitatively it looks as follows.

\medskip

\paragraph{`How much contextuality'?} So far, only qualitative ideas were provided.
The next step is to try to evaluate them, putting the question
`How much contextuality'? A possible transparent answer was
recently proposed in \cite{svozil2011}. We take a representative
sampling of observables, and simply check the ratio of the
triples, for which Accardi inequalities \eqref{eaccardi} are
violated.

Using the ideas of \cite{svozil2011}, the rate of personalization can be evaluated in a similar way.
First, by random sampling, triples of properties, that is, yes-no queries are picked. Then, for each triple, the
transition probability matrices \eqref{ebistoch} are calculated. For each particular sample triple the inequalities
\eqref{eaccardi} are checked. Then the ratio of samples is calculated:
\begin{equation}\label{ekarleval}
    {\rm \textbf{Pers}}=\frac{\mbox{number of triples violating \eqref{eaccardi}}}{\mbox{total number of sampled triples}}
\end{equation}

\section*{Conclusions}

Vector models of IR become more and more popular, first of all
because they make it possible to carry out multi-document actions.
In this paper I dwell on a QIA framework \cite{piwowarski2010}.
The basic ingredient of QIA framework is a Hilbert space HH called
the information need space. In it simplest form, IN space is
linear space of elementary (atomic) topics. In my approach, I
suggest to start introducing the IN space to satisfy the necessary
amount of capturing contextuality. The ideology of IN space is the
closest to that of quantum mechanics. In QM, the state space of a
system is a space of some internal (in the deepest possible sense)
features of a system, while the observables are expressed in terms
of operators and other derived structures on the state space.
Similar things happen in QIA approach. The pace of information
needs exists per se, we may treat it as spanned on elementary
entities, but this will be nothing, but a representation of this
space. The source of emergence of this space lies in the
multicontextual structure described in the previous section.
Furthermore, as pointed in \cite{liang-specker2011},
\cite{melucci2012}, the correlations, which occur in IR environment may
even be stronger than quantum ones. In this case a straightforward
Hilbert space model may fail to work properly, and we may call
'foil quantum theories' to grip these situations.

\medskip

So far, I was interested in information retrieval situations, when
the result of a particular action may depend on other actions,
which the IR agent could in principle do alongside with the actions actually performed.
This phenomenon is called contextuality, we encounter it
in IR, we have to take it into account, to work with it. A similar
kind of dependence takes place in quantum mechanics.

\medskip

\begin{tabular}{l|c|l}
  Quantum Mechanics &  & Information Retrieval \\
   & $\leftrightarrow$ &  \\
  contextuality &  & personalization \\
\end{tabular}

\medskip

\noindent The difference is that in QM contextuality appears by itself, not
being originated by some 'internal mechanisms'. The situations
where contextuality occurs depend on the state space of the
system the structure of observables involved. In the realm of QM we
can quantitatively evaluate the rate of contextuality
\cite{svozil2011}. The origin of contextuality effects in IR stems from
personalization of query scenarios. The personalization, in turn,
can be quantified by memory resources required to keep tracking
the information needs of a particular user (note that `user' in
this context might not be a single person, nor even a `person' at
all). The idea of this Note was to demonstrate that using quantum mechanics formalism, we can quantify the rate of
personalization in particular IR environments. To do this, I suggest to
reverse the procedure of estimation of memory cost of quantum
contextuality based on simulating quantum systems by finite
automata. Instead, a KR scenario (which as a matter of fact is a
sequence of queries upon a finite automaton) is suggested to be
simulated by appropriate scenario of quantum measurement,
demonstrating the same contextuality features. As a result, a
Hilbert space of appropriate quantum system emerges together with
a collection of observables. This Hilbert space is suggested to
play the role of information need space, which is developed within
QIA (quantum information access) framework for Information
Retrieval. Technically, the IN space is built starting from
Greechie-like digramss (pasted overlapping contexts, see
Section \ref{sgreechie} above, capturing the particular IR environment.
QIA framework provides more flexible machinery to deal with
information needs than any classical probabilistic approach by
that simple reason that it incorporates the latter. But we should
be aware that it is not ultimately general. In quantum realm, we
have non-classical correlations and the present state of our
knowledge shows that quantum mechanics is enough to explain all
them. However, IR may in principle provide stronger-than-quantum
correlations. For them, `foils of quantum theory' - the
operational theories, which do not compete with quantum mechanics,
but generalize it to the extent not demanded in modern physics
\cite{liang-specker2011}, these theories may be of help in Information
Processing.

\paragraph{Acknowledgments.} I greatly appreciate Cris Calude,
 Karl Svozil and Jozef Tkadlec for stimulating discussions on
 quantum contextuality during my stay in Technical University of
 Vienna, supported by the Ausseninstitut and the Institute of
 Theoretical Physics of the Vienna University of Technology.
 A financial support from Russian Basic Research Foundation (grant
10-06-00178a) is appreciated.


\end{document}